\documentclass[showpacs,preprintnumbers,amsmath,amssymb,prl,aps,onecolumn]{revtex4}
\usepackage{graphicx}
\usepackage{bm}
\usepackage[usenames,dvipsnames]{color} 
\usepackage{ulem}
\usepackage{amsmath}

\newcommand{\bk}{{\bf k}} 
\newcommand{\bq}{{\bf q}}
 
\newcommand{\bp}{{\bf p}}

\makeatletter
\let\columncolor\relax
\def\endarray{%
   \adl@endarray \egroup \adl@arrayrestore \egroup
   \gdef\@preamble{}\CT@end}
\def\@array{%
   \adl@everyvbox\everyvbox
   \everyvbox{\adl@arrayinit \the\adl@everyvbox \everyvbox\adl@everyvbox}%
   \global\let\adl@hdashline@bgcolorsaved\adl@hdashline@bgcolor
   \global\let\adl@hdashline@bgcolor\@empty
   \adl@array}
\let\@@array\@array
\expandafter\def\expandafter\adl@arrayrestore\expandafter{%
   \global\let\adl@hdashline@bgcolor\adl@hdashline@bgcolorsaved}
\global\let\adl@hdashline@bgcolorsaved\@empty
\global\let\adl@hdashline@bgcolor\@empty
\def\adl@@colhtdp{%
   \ifdim\adl@height<\@tempdima \global\adl@height\@tempdima \fi
   \ifdim\adl@depth<\dp\z@  \global\adl@depth\dp\z@  \fi}
\def\adl@ihdashline[#1/#2]{%
   \adl@hline\adl@connect\arrayrulewidth
   \ifnum0=`{\fi}%
   \adl@hdashline@bgcolor\crcr\noalign{\vskip-\arrayrulewidth}%
   \multispan{\adl@columns}\unskip \adl@hcline\z@[#1/#2]%
   \noalign{\ifnum0=`}\fi
   \futurelet\@tempa\adl@xhline}
\def\@classz{%
   \@classx
   \@tempcnta\count@
   \prepnext@tok
   \expandafter\CT@extract\the\toks\@tempcnta\columncolor!\@nil
   \@addtopreamble{%
      \setbox\z@\hbox\bgroup\bgroup
         \ifcase\@chnum
            \hskip\stretch{.5}\kern\z@
            \d@llarbegin \insert@column \d@llarend
            \hskip\stretch{.5}%
         \or \d@llarbegin \insert@column \d@llarend \hfill
         \or \hfill\kern\z@ \d@llarbegin \insert@column \d@llarend
         \or $\vcenter \@startpbox{\@nextchar}\insert@column \@endpbox$%
         \or \vtop \@startpbox{\@nextchar}\insert@column \@endpbox
         \or \vbox \@startpbox{\@nextchar}\insert@column \@endpbox
         \fi
      \egroup\egroup
      \begingroup
      \CT@setup
      \CT@column@color
      \CT@row@color
      \CT@do@color
      \endgroup
      \@tempdima\ht\z@
      \advance\@tempdima\minrowclearance
      \adl@colhtdp
      \vrule\@height\@tempdima\@width\z@
      \unhbox\z@}%
   \adl@setup@bgcolor
   \prepnext@tok}
\def\adl@setup@bgcolor{%
   \xdef\adl@hdashline@bgcolor{%
      \adl@hdashline@bgcolor
      \ifx\adl@hdashline@bgcolor\@empty\else &\fi
      \omit
      \setbox\z@\hbox{}\ht\z@\arrayrulewidth
      \begingroup
      \CT@setup
      \hskip\@tempdimb
      \CT@column@color
      \CT@do@color\unskip
      \hskip\@tempdimc
      \endgroup
      \box\z@}}%
\def\multicolumn#1#2#3{%
   \multispan{#1}%
   \begingroup\begingroup
   \def\adl@arrayrule{\adl@mcarrayrule{#1}}%
   \def\adl@arraydashrule{\adl@mcarraydashrule{#1}}%
   \def\adl@argarraydashrule{\adl@mcargarraydashrule{#1}}%
   \let\@addamp\adl@mcaddamp
   \let\adl@setup@bgcolor\relax
   \@mkpream{#2}\@addtopreamble\@empty
   \global\let\adl@preamble\@preamble\endgroup
   \let\@preamble\adl@preamble
   \def\@sharp{#3}\let\protect\relax
   \adl@activatepbox
   \adl@preaminit
   \let\CT@row@color\relaxx

   \let\CT@column@color\relax
   \let\CT@do@color\relax
   \@arstrut \@preamble\hbox{}\endgroup
   \global\advance\adl@currentcolumn#1\ignorespaces}

\begin{document}
\title{Hybridization of Bogoliubov-quasiparticles between adjacent CuO$_2$ layers in the triple-layer cuprate Bi$_2$Sr$_2$Ca$_2$Cu$_3$O$_{10+\delta}$ studied by ARPES}

\renewcommand{\theequation}{S\arabic{equation}}

\author{S.~Ideta$^{1,2*}$, S.~Johnston$^3$, T.~Yoshida$^4$, K.~Tanaka$^2$, M.~Mori$^5$, H.~Anzai$^{6,7}$, A.~Ino$^{7,8,9}$, M.~Arita$^8$, H.~ Namatame$^8$, M.~Taniguchi$^{7,8}$, S.~Ishida$^{1,12}$, K.~Takashima$^1$, K.~M.~Kojima$^{1,10,13}$, T.~P.~Devereaux$^{11}$, S.~Uchida$^{1,12}$, and A.~Fujimori$^{1,14}$ }

\affiliation{
$^1$ Department of Physics, University of Tokyo, Bunkyo-ku, Tokyo 113-0033, Japan\\
$^2$ Institute for Molecular Science, UVSOR-III Synchrotron, Okazaki 444-8585, Japan\\
$^3$ Department of Physics and Astronomy, The University of Tennessee, Knoxville, TN 37996, USA\\
$^4$ Department of Human and Environmental studies, Kyoto University, Sakyo-ku, Kyoto 606-8501, Japan\\
$^5$ Advanced Science Research Center, Japan Atomic Energy Agency, Tokai 319-1195, Japan\\ 
$^6$ Graduate School of Engineering, Osaka Prefecture University, Sakai 599-8531, Japan\\
$^7$ Graduate School of Science, Hiroshima University, Higashi-Hiroshima 739-8526, Japan\\
$^8$ Hiroshima Synchrotron Radiation Center, Hiroshima University, Higashi-Hiroshima 739-0046, Japan\\
$^{9}$ Department of Education and Creation Engineering, Kurume Institute of Technology, Fukuoka 2286-66, Japan\\
$^{10}$ J-PARC Center and Institute of Materials Structure Science, KEK, Tsukuba, Ibaraki 305-0801, Japan\\
$^{11}$ Geballe Laboratory for Advanced Materials, Stanford University, Stanford, CA4305, USA\\
$^{12}$  National Institute of Advanced Industrial Science and Technology, Tsukuba, Ibaraki 305-8568, Japan\\
$^{13}$ Centre for Molecular and Materials Science, TRIUMF, 4004 Vancouver, Canada\\
$^{14}$ Department of Applied Physics, Waseda University, Shinjuku-ku, Tokyo 169-8555, Japan 
}

\date{\today}
\begin{abstract}
Hybridization of Bogoliubov quasiparticles (BQPs) between the CuO$_2$ layers in the triple-layer cuprate high-temperature superconductor Bi$_2$Sr$_2$Ca$_2$Cu$_3$O$_{10+\delta}$ is studied by angle-resolved photoemission spectroscopy (ARPES). In the superconducting state, an anti-crossing gap opens between the outer- and inner-BQP bands, which we attribute primarily to interlayer single-particle hopping with possible contributions from interlayer Cooper pairing.  We find that the $d$-wave superconducting gap of both BQP bands smoothly develops with momentum without abrupt jump in contrast to a previous ARPES study. Hybridization between the BQPs also gradually increases in going from the off-nodal to the anti-nodal region, which is explained by the momentum-dependence of the interlayer single-particle hopping. As possible mechanisms for the enhancement of the superconducting transition temperature, the hybridization between the BQPs as well as the combination of phonon modes of the triple CuO$_2$ layers and spin fluctuations represented by a four-well model are discussed. 
\end{abstract}

\maketitle 
In the cuprate high-temperature superconductors (HTSCs), the microscopic origin of the high superconducting (SC) transition temperatures ($T_{\rm{c}}$) has been a major issue since their discovery. Among accumulated experimental results, the relationship between the $T_{\rm{c}}$ and the number of adjacent CuO$_2$ layers ($n$) has attracted strong interest: The $T_{\rm{c}}$ increases from $n$ = 1 to $n$ = 2, reaches a maximum at $n$ = 3, and then decreases for $n$ $>$ 3 [1, 2]. A lot of theoretical [3-15] and experimental [16, 17] effort has been made to elucidate the mechanism for the non-monotonic behavior of $T_{\rm{c}}$. Both the hopping of single particles and that of Cooper pairs between the CuO$_2$ layers have been proposed to play an essential role in boosting the $T_{\rm{c}}$ in the multi-layer cuprates, but experimental information about the interlayer hopping in the SC state is lacking.  

So far there has been a consensus that the single-particle hopping in the normal state is reflected in the angle-resolved photoemission spectroscopy (ARPES) spectra of double-layer [18] and triple-layer [19, 20] cuprates. When Cooper pairs are formed below $T_{\rm{c}}$, one can directly observe the opening of a SC gap $\Delta(\bm{k})$ at the Fermi level ($E_{\rm{F}}$) by ARPES [21]. That is, by ARPES measurements, one can directly prove the Cooper pairing through the observation of Bogoliubov quasiparticles (BQPs) resulting from hybridization between the electron and hole branches at $E_{\rm{F}}$ [21-24]. Recently, a laser ARPES study of the optimally doped triple-layer HTSC Bi$_2$Sr$_2$Ca$_2$Cu$_3$O$_{10+\delta}$ (Bi2223) by Kunisada $et$ $al$. [25] has shown that in the off-nodal region, where the BQP band of the outer CuO$_2$ plane (OP) of hole character and that of the inner CuO$_2$ plane (IP) of electron character cross each other below $E_{\rm{F}}$, the OP and IP are hybridized with each other and open a gap at the crossing point (an anti-crossing gap between the two BQP bands). Their study has also shown that, as one moves from the nodal to the off-nodal regions, the SC gap of the IP band shows a sudden increase. The authors have suggested that this sudden increase of the SC gap of the IP band plays an important role in achieving the enhancement of $T_{\rm{c}}$ in Bi2223 [25].

In this Letter, we have re-investigated the BQP bands in the optimally-doped Bi2223 and examined the hybridization between the BQP bands of the OP and IP. We have reproduced the results of Kunisada $et$ $al$. [25] that the new energy gap opens at the crossing point of the BQP band of the OP and that of the IP, and that the anti-crossing gap closes above $T_{\rm{c}}$. However, in contrast to their results [25], the momentum dependence of the SC gap magnitude is smooth without any sudden increase in going from the node to the anti-node. Furthermore, we have observed that the anti-crossing gap is strongly momentum dependent and is proportional to the square of the $d$-wave SC order parameter. We attribute the anti-crossing gap to the hybridization of the BQP bands of OP and IP through interlayer single-particle hopping but a finite contribution from interlayer Cooper-pair hopping may also be present [14]. 

High-quality single crystals of optimally doped Bi2223 ($T_{\rm{c}}$ = 110 K) were grown by the traveling solvent floating zone method. See also Sec. I of Supplemental Material (SM) [29]. ARPES experiments were carried out at BL9A of Hiroshima Synchrotron Radiation Center and BL7U of UVSOR-III synchrotron. Samples were cleaved $in$ $situ$ in an ultrahigh vacuum of $\sim$5$\times$10$^{-9}$ Pa at $T$ = 10 K. Measurements were performed with photon energies of $h\nu$ = 7.54-11.95 eV. The total energy resolution was $\Delta{E}\sim$ 3-7 meV. 

Figure 1 shows ARPES intensity plots taken at several photon energies in the SC state ($T$ = 10 K). The relative spectral intensities of the OP and IP bands change with photon energy due to matrix-element effect [19]. Since in synchrotron radiation ARPES, one can continuously change the photon energy, the electronic structure can be measured with varying matrix-element effects in contrast to laser ARPES, where the photon energy is fixed. As a result, one can selectively enhance the IP or OP band, as shown in a previous study [19]. Synchrotron radiation ARPES can also suppress the space charge effect compared to laser ARPES [26]. The SC gap of the OP band ($\Delta_{\rm{OP}}$) and that of the IP band ($\Delta_{\rm{IP}}$) increase in going from cut $\#$1 to cut $\#$4 following the momentum dependence of the $d$-wave order parameter $\Delta_{\rm{OP}}$ = $\Delta_{\rm{OP}}^0$[$\cos$($k_xa)-\cos$($k_ya$)]/2 and $\Delta_{\rm{IP}}$ =$\Delta_{\rm{IP}}^0$[$\cos$($k_xa)-\cos(k_ya)$]/2 [19]. Here, $\Delta_{\rm{IP}}^0$ is significantly larger than $\Delta_{\rm{OP}}^0$ as reported previously [19, 20]. In BCS theory, when a SC gap opens, the band dispersion shows a back-bending due to particle-hole hybridization, as has been observed by ARPES studies of HTSCs such as Bi$_2$Sr$_2$CaCu$_2$O$_{8+\delta}$ (Bi2212) and Bi$_2$Sr$_2$CuO$_{4+\delta}$ [21]. In Bi2223, too, the ARPES intensity plot (Fig. 1) clearly shows a back bending both for the OP and IP bands. In addition, we have observed an anomaly which becomes more pronounced in going from cut $\#$3 to cut $\#$4 as indicated by arrows. 

To examine the dispersion anomaly of the IP band in more detail, the ARPES intensity plot in the vicinity (within $\sim$70 meV) of $E_{\rm{F}}$ and the corresponding energy-distribution curves (EDCs) are shown in Figs. 2(a)-2(d) and 2(e)-2(h), respectively, on an expanded energy scale. As the BQP band approaches $E_{\rm{F}}$, the dispersion flattens, bends back around the Fermi momentum ($k_{\rm{F}}$), and the spectral weight is transferred from the electron branch to the hole branch, which is characteristic of BQPs [21]. As one moves away from the node and $\Delta_{\rm{OP}}$ and $\Delta_{\rm{IP}}$ increase, the hole branch of the OP band gradually overlaps the electron branch of the IP band and they cross each other. Around the crossing point, EDCs [Figs. 2(e)-2(h)] show a double peak structure indicating the opening of an anti-crossing gap. Here, the processes of the data analysis are as follows: The peak positions of the EDCs for the OP and IP bands are marked by open and filled circles, respectively, where the EDC peak positions are determined by fitted results using a phenomenological spectral function convoluted with the total energy resolution [35]. The red dotted curves in Figs. 2(a)-2(d) are fitted results using Eq. (S7) in Sec. IV of SM. The SC gaps ($\Delta_{\rm{OP}}$ and $\Delta_{\rm{IP}}$) estimated from this analysis were used to derive relevant parameters in Eqs. (S7)-(S9) of SM. Here, the fitted results are the same when Eq. (S9) is used instead of Eq. (S7) to fit the ARPES data. The EDCs showing the anti-crossing gaps in Figs. 2(e)-2(h) are collected in Fig. 2(i). Here, in Bi2212, bonding and anti-bonding bands do not hybridize due to the different orbital symmetries. The magnitude of the anti-crossing gap gradually increases as one goes away from the node towards the anti-node. We denote the anti-crossing gap magnitude by 2$\Delta_{t_{\perp}}(\bm{k})$ hereafter because it is proportional to the single-particle interlayer hopping $t_{\perp}(\bm{k})$ according to Eq. (S8) in Sec. IV of SM. In Fig. 2(j), the $\Delta_{t_{\perp}}(\bm{k})$ values estimated from the present data and energy gaps ($\Delta_{\rm{OP}}$, $\Delta_{\rm{IP}}$) are plotted against the $d$-wave order parameter. The plot shows that $\Delta_{t_{\perp}}(\bm{k})$ increases from the node towards the anti-node as a quadratic function of the $d$-wave order parameter. In the previous ARPES study by Kunisada $et$ $al$. [25], the momentum dependence of $\Delta_{t_{\perp}}(\bm{k})$ was ignored and $\Delta_{\rm{IP}}(\bm{k})$ was considered to suddenly increase around $|\cos(k_xa)-\cos(k_ya)|$/2 $\sim$ 0.4, where our $\Delta_{t_{\perp}}(\bm{k})$ significantly increases. That is, they considered that, beyond this momentum, $\Delta_{\rm{IP}}(\bm{k})$ was deviated from the $d$-wave gap and enhanced as plotted in Fig. 2(j) (see also Sec. VI of Fig. S4 in SM which shows AREPS data near the anti-nodal region).

Now, we analyze the present experimental data using the tight-binding model for coupled CuO$_2$ planes. We assume that the OP and IP bands are coupled primarily through the single-particle hopping between the OP and IP to be $t_{\perp}(\bm{k})$ = $t_{\perp}^{(0)}$+$t_{\perp}^{(1)}$[$\cos(k_xa)-\cos(k_ya)$]$^2$/4, where $t_{\perp}^{(0)}$ = 0 meV and $t_{\perp}^{(1)}$= 56 meV [5, 27]. In Bi2223, the energy band of the two OPs are originally identical and form bonding and anti-bonding combinations, and the IP band is hybridized with the bonding combinations of the OP bands. The potential difference between the OP and IP, $\epsilon_{\rm{OP}}-\epsilon_{\rm{IP}}$ = 65 meV, can be measured from the OP-IP band splitting in the nodal direction, where $t_{\perp}(\bm{k}$) = 0 [28]. The BQP band of the bonding OP and that of the IP can be hybridized with each other through $t_{\perp}(\bm{k})$. Then 2$\Delta_{t_{\perp}}$($\bm{k}$) can be expressed as Eq. (S8) as derived in Sec. IV of SM. Then the experimental 2$\Delta_{t_{\perp}}$($\bm{k}$) can be fitted by Eq. (S8) as shown by a black curve in Fig. 2(j). It should be noted that all the momentum-dependent parameters $\Delta_{\rm{IP}}$($\bm{k}$), $\Delta_{\rm{OP}}$($\bm{k}$), and $\Delta_{t_{\perp}}$($\bm{k}$) which describe the effect of superconductivity are simple, smooth functions of $\bm{k}$ and that no abrupt change around cuts $\#$3 and $\#$4 as claimed by the previous ARPES work [25] is seen in our analysis. The abrupt change of $\Delta_{\rm{IP}}(\bm{k})$ reported in Ref. [25] would be due to their use of energy positions $after$ the OP-IP hybridization, while we define $\Delta_{\rm{IP}}$($\bm{k}$) using energy positions $before$ the hybridization.

In order to simulate the present ARPES spectra, we have performed a model calculation for the normal state and the SC state of Bi2223 by using the interlayer single-particle hopping parameter $t_{\perp}^{(1)}$ = 56 meV (for details, see Sec. VII of SM). Calculated spectral function are displayed in Figs. 3(a1)-3(a4), corresponding to the experimental data in Fig. 1. In order to reproduce the high energy kinks for the OP ($E_{\rm{kink}}$ $\sim$ 80 meV) and IP ($E_{\rm{kink}}$ $\sim$ 100 meV) bands of the measured ARPES spectra [30].  Here, the kink energy of Bi2223 in the IP band is much larger than that of Bi2212 in the nodal and off-nodal directions (see Sec. II of SM). Taking into account the contribution of acoustic phonons, the flatness of the top of the IP band (cuts $\#$3 - $\#$4 of Figs. 1 and 2) seen in the experiment is reproduced to some extent. Fermi surfaces calculated using $t_{\perp}^{(1)}$ = 56 meV are also shown in Fig. 3(b) and are in good agreement with experiment. The calculation also reproduces the dispersion anomalies of the IP band as due to the hybridization between the hole branch of the OP band and the electron branch of the IP band. Thus, our model calculation assuming single-particle hopping can reproduce the hybridization between the OP and IP BQP bands. Note that the present experimental and theoretical studies do not assume any sudden increase of the SC gap of IP band, which is reported to occur near cuts $\#$3 and $\#$4 in Ref. [25] [see Fig. 2(j)].

The most intriguing and important question relevant to the present study is how much the interlayer single-particle hopping and the obtained $\Delta_{t_{\perp}}(\bm{k})$ contribute to the enhancement of $T_{\rm{c}}$ in Bi2223. Firstly, we note that ARPES cannot distinguish between the effect of $\Delta_{t_{\perp}}$ and the effect of interlayer Cooper-pair hopping, the matrix element of which we denote by $\Delta_{\rm{OP-IP}}$, (see Eq. (S9) in Sec. V of SM) because both lead to the same ARPES spectra as shown in Sec. IV-VI of SM. In fact, 2$\Delta_{t_{\perp}}$ for the BQP bands of OP and IP can also be reproduced in the Cooper-pair hopping scenario, as described in Sec. V of SM. Therefore, we cannot exclude the interlayer Cooper pairing as a mechanism to increase the $T_{\rm{c}}$ in Bi2223 [3, 4, 8-10, 14]. Theoretical studies which take into account pair hopping have shown that the $T_{\rm{c}}$ of a double-layer system is enhanced by a factor of three compared with the $T_{\rm{c}}$ without interlayer pair hopping but, one needs to assume a too large pair hopping matrix element to reproduce the $T_{\rm{c}}$ enhancement observed by experiment [14]. According to another theoretical study, using cluster dynamical mean-filed theory, the combination of underdoped and overdoped CuO$_2$ layers shows an increase of the SC order parameter of the overdoped CuO$_2$ layer and $T_{\rm{c}}$ through proximity effect [13]. According to this mechanism, however, interlayer hopping as large as $t_{\perp}$= 0.5$t$, where $t$ is the in-plane nearest neighbor Cu-Cu transfer integral, is necessary to enhance the $T_{\rm{c}}$ by 6$\%$ compared with its value in the uniform system. A composite model system consisting a layer with a large pairing energy scale ($\Delta$ $>$ 0) but small superfluid density and a metallic layer ($\Delta$ = 0) with high superfluid density via single-particle hopping, where $T_{\rm{c}}$ = 0 for both layers, also gives rise to a high $T_{\rm{c}}$ of $\sim$$\Delta$/2 under the optimal magnitude of $t_{\perp}$ [32, 34]. 

In order to discuss a possible mechanism for the $T_{\rm{c}}$ enhancement in triple-layer cuprates, a simple four-well model calculation is performed assuming contributions from four bosonic modes, namely, the acoustic (Ac) phonon, the $c$-axis buckling ($B_{1g}$) phonon, the in-plane breathing (Br) phonon, and spin fluctuations (SFs) (for details, see Sec. IX of SM and Ref. [33]). $T_{\rm{c}}$ is roughly estimated and plotted as a function of the Ac phonon coupling strength $\lambda_{\phi}$ for pairing in Fig. 4. According to the calculation, the $T_{\rm{c}}$ is generally enhanced compared to the three-well model calculation reported in [33], which is relevant to double-layer cuprates. Contributions from the individual $B_{1g}$ phonon modes of OP and IP combined with SFs lead to $d$-wave pairing, whereas the Br phonons lead to pair-breaking. As reported in the previous study [30], coupling with the $B_{1g}$ phonon mode is important to reproduce the strong kink structure in Bi2223 and may play an important role to increase $T_{\rm{c}}$. As shown in Fig. 4(b), the present study shows that coupling with the $B_{1g}$ phonon indeed contributes to the $T_{\rm{c}}$ enhancement. The present four-well model calculation suggests that electron-phonon coupling alone may not be sufficient to cause the observed $T_{\rm{c}}$ in the triple-layer cuprate, and the combination of spin fluctuations and phonons is essential. The CuO$_2$ planes in the double- and triple-layer cuprates are placed in a strong electric field along the $c$ axis, making electron-phonon coupling in the double- and triple-layer cuprates stronger than that in the single-layer cuprates. The increased degree of symmetry breaking, namely, the increased strength of electric field in the triple-layer cuprates leads to the enhancement of electron-phonon coupling [31]. The large difference in the hole concentrations of IP and OP in Bi2223 also contributes to the local electric field and the degree of symmetry breaking, resulting in the stronger electron-phonon coupling in Bi2223 than in Bi2212 and stronger enhancement of the $T_{\rm{c}}$ in the triple-layer cuprate. Hence, the electron-phonon interaction plays an importance role, and the $T_{\rm{c}}$ enhancement mechanism should be clarified along this line in future work.

In conclusion, we have performed ARPES experiments on Bi$_2$Sr$_2$Ca$_2$Cu$_3$O$_{10+\delta}$, and observed a signature of the hybridization of BQPs between OP and IP. Due to the hybridization, a new gap opens at the crossing point of the OP and IP BQP bands, and is explained by the interlayer single-particle hopping of $t_{\perp}^{(1)} \sim $56 meV. As another candidate of the anti-crossing gap, the effect of interlayer Cooper-pair hopping could not be isolated from the present data, but might be important to increase the $T_{\rm{c}}$ of Bi2223. A mechanism of the $T_{\rm{c}}$ enhancement is also suggested using the four-well model that a combination of phonon modes of OP and IP and spin fluctuations is included. As for the question of how interlayer interactions, i.e., single-particle hopping versus Cooper pair hopping, contribute to the enhancement of superconductivity, further systematic studies are needed both experimentally and theoretically. 

We would like to acknowledge discussions with T. Morinari and experimental supports from M. Hashimoto. ARPES experiments were carried out at Hiroshima Synchrotron Radiation Center, Hiroshima University (Proposal No. 07-A-10 and No. 08-A-35) and UVSOR-III synchrotron (Proposal No. 27-813, No. 28-549, and No. 29-549). This work was supported by Grants-in-Aid for Scientific Research (15K17709, 22740221, 15H02109, 16K05445, and 19K03741) from JSPS, “Program for Promoting Researches on the Supercomputer Fugaku” (Basic Science for Emergence and Functionality in Quantum Matter) from MEXT, a Research Fellowship for Young Scientists from JSPJ, and Murata Science Foundation.

{$^{*}$Present address: Hiroshima Synchrotron Radiation Center, Hiroshima University, Higashi-Hiroshima, 739-0046, Japan}

\begin{figure} [h]
 \includegraphics[width=10cm]{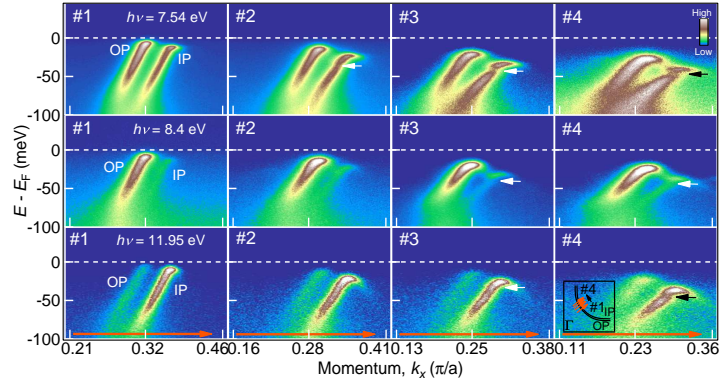}
 \caption{\label{Fig:1} ARPES spectra of Bi2223 in the superconducting (SC) state ($T$ = 10 K) taken using circularly polarized light. Energy-momentum ($E$-$k$) intensity plot is shown for the outer and inner CuO$_2$ planes (OP and IP) in the off-nodal region taken using three photon energies. The relative intensities of the OP and IP change with photon energy. Orange arrows in the inset show cuts $\#$1 - $\#$4 along which the intensity plots are displayed. A dispersion anomaly in the IP band is indicated by arrows.}
\end{figure}
\begin{figure} [h]
 \includegraphics[width=16cm]{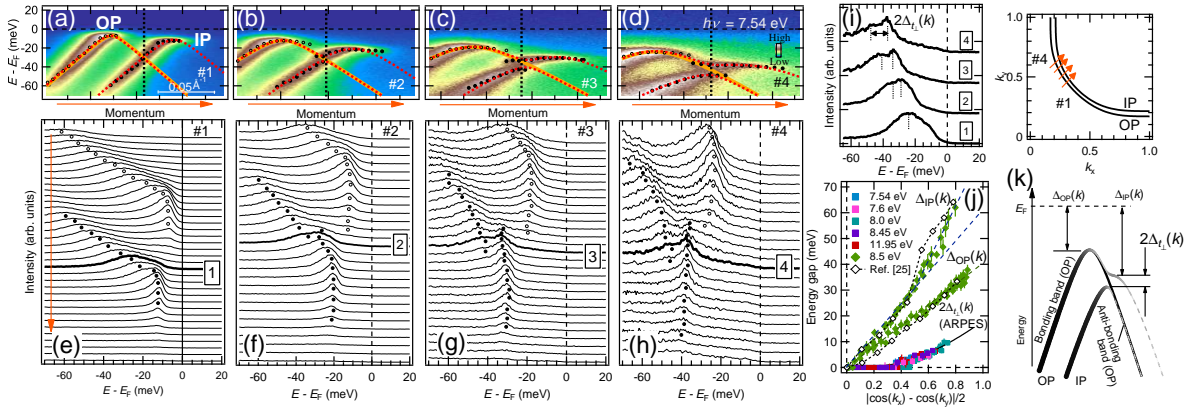}
 \caption{\label{Fig:2} ARPES spectra of the OP and IP bands in Bi2223 taken using circular-polarized light. (a)-(d): $E$-$k$ plots near $E_{\rm{F}}$. The measured momentum cuts are shown in the right panel. (e)-(h): Energy distribution curves (EDCs) corresponding to (a)-(d). Peaks of the EDCs for the OP (IP) band are marked by open (filled) circles. These markers are put on panels (a)-(d), too. Dotted curves in panels (a)-(d) are fitting results using Eq.  (S7) of Sec. IV of SM [29]. Fitting parameters $\Delta_{\rm{OP}}$ and $\Delta_{\rm{IP}}$ are plotted in panel (j), and $\Delta_{t_{\perp}}$ are calculated using Eq. (S8) with these $\Delta_{\rm{OP}}$ and $\Delta_{\rm{IP}}$ values and $t_{\perp}^{(1)}$ = 56 meV. The orange solid curve is the anti-bonding OP band, which does not hybridize with the IP band and does not show an anti-crossing gap. (i): EDCs at momenta where the hole branch of the OP band and the electron branch of the IP band cross extracted from panels (e)-(h). Vertical lines show peak positions of the EDCs. The energy difference between two peaks defines 2$\Delta_{t_{\perp}}(\bm{k}$). (j) Momentum dependence of $\Delta_{t_{\perp}}(\bm{k})$ plotted against the $d$-wave order parameter. $\Delta_{\rm{OP}}(\bm{k})$ and $\Delta_{\rm{IP}}$($\bm{k}$) ($\propto$$|\cos(k_xa)-\cos(k_ya)|$/2) are obtained from the present ARPES study on Bi2223 and the black curve is a fit using $t_{\perp}^{(1)}$= 56 meV in Eq. (S8). SC gaps taken from Ref. [25] are also plotted using white diamonds for both the OP and IP bands. (k) Schematic illustration of the hybridization between the OP and IP BQP bands in the off-nodal region. Spectral weight of the BQP bands for the OP (bonding band) and IP are indicated by the thickness of the curve. Anti-bonding OP band is shown by a black solid curve. }

\end{figure}
\begin{figure} [t]
 \includegraphics[width=16cm]{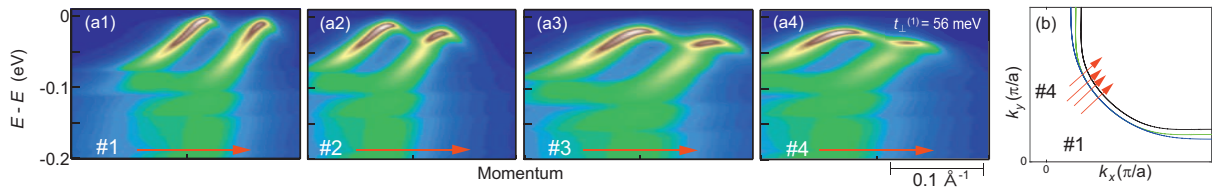}
 \caption{\label{Fig:3} Calculated spectral function for the SC state of Bi2223. Coupling to the out-of-plane bond-backing (${B}_{1g}$) and in-plane bond-stretching (breathing) optical phonons with energies of 36 and 70 meV, respectively, and acoustic phonons with momentum transfer $q_{\rm{TF}}$ = 0.6/$a$ for the outer planes and $q_{\rm{TF}}$ = 0.4/$a$ for the inner CuO$_2$ plane are taken into account. See also Sec. VII of SM [29]. (a) Intensity mapping along four cuts $\#$1-$\#$4 from the node (a1) to the off-node (a4) shown in panel (b). Interlayer hopping $t_{\perp}^{(1)}$ of 56 meV is assumed. The FSs in panels (b) are calculated using $t_{\perp}^{(1)}$ = 56 meV. $\Delta_{t_{\perp}}$ used in the simulation is taken from the experimental data.}
\end{figure}
\begin{figure} [h]
 \includegraphics[width=16cm]{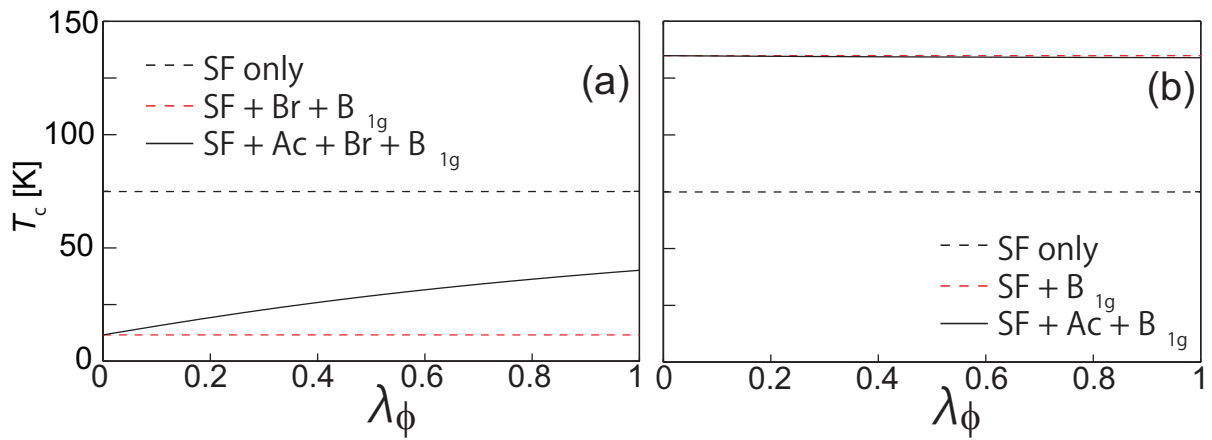}
 \caption{\label{Fig:4}  $T_{\rm{c}}$ calculated using a four-well model for Bi2223 as a function of the acoustic (Ac) phonon coupling strength $\lambda_{\phi}$. The $\lambda_{\phi}$ for the spin fluctuation (SF) is set to provide a base $T_{\rm{c}}$ of 75 K. While the addition of the $B_{1g}$ and Ac phonon modes raises the $T_{\rm{c}}$, the Br phonon mode contributes to the pair breaking. The addition of the $B_{\rm{1g}}$ phonon mode dramatically raises the $T_{\rm{c}}$. See also Sec. IX of SM [29] for more detail.}
\end{figure}
\clearpage

\renewcommand{\thefigure}{S\arabic{figure}}
\setcounter{figure}{0}
\newpage
\subsection{\Large{Supplemental Material for "Hybridization of Bogoliubov-quasiparticles between adjacent CuO$_2$ layers in the triple-layer cuprate Bi$_2$Sr$_2$Ca$_2$Cu$_3$O$_{10+\delta}"$ studied by ARPES}}

\subsection{\Large{I. Superconducting transition temperature of samples for ARPES}}
The superconducting transition temperature ($T_{\rm{c}}$) of optimally doped Bi2223 was measured by SQUID experiment. The onset of $T_{\rm{c}}$ of 110 K is estimated as shown in Fig. S1. 
\begin{figure}[h] 
 \includegraphics[width=5cm]{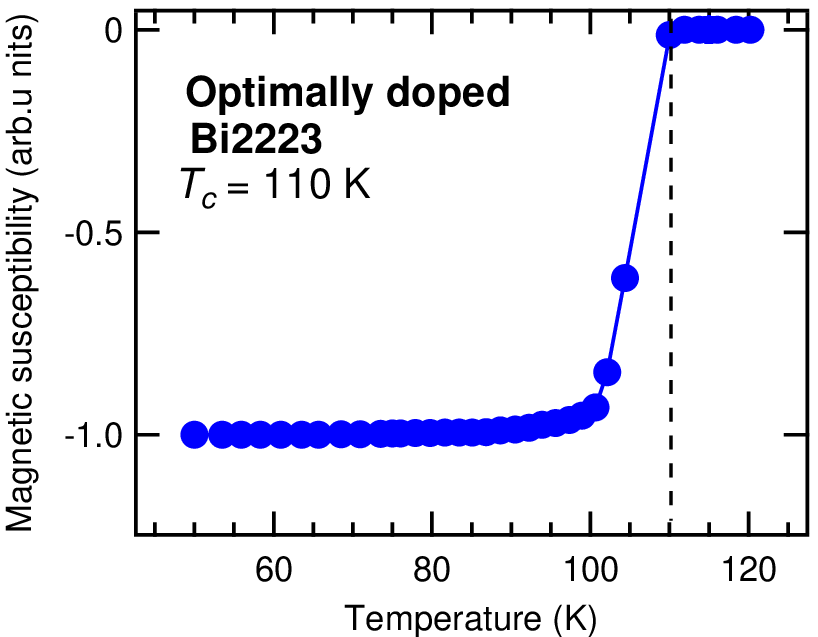}
 \caption{\label{Fig:S1}Normalized magnetic susceptibility of the single crystal Bi2223 measured by the present ARPES experiments.}
\end{figure}
\subsection{\Large{II. Large kink energy in the IP band of Bi2223}}
\begin{figure}[h] 
 \includegraphics[width=17cm]{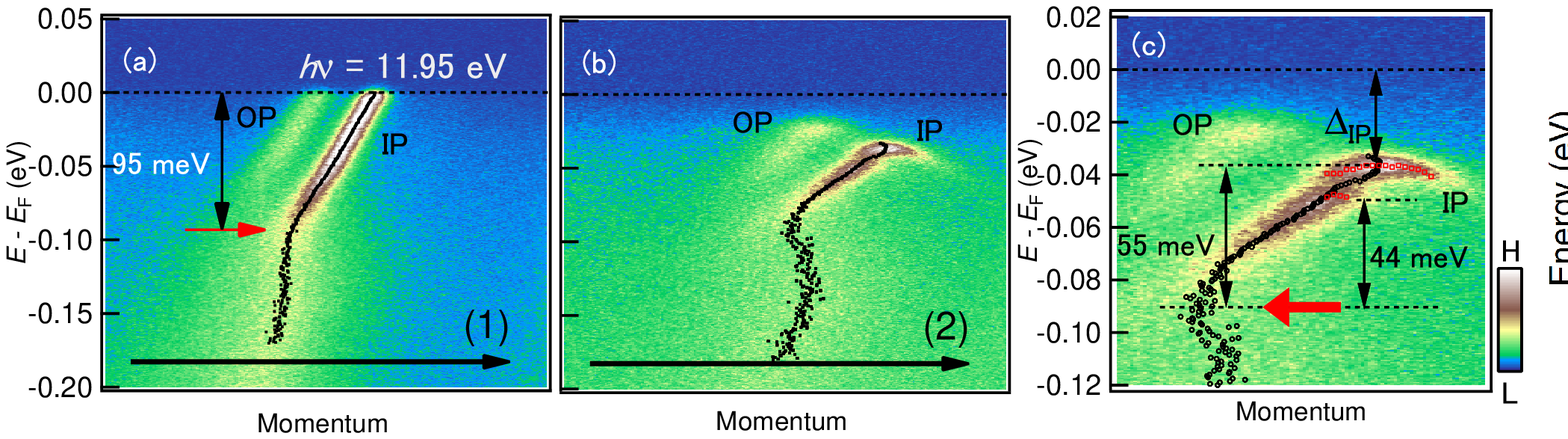}
 \caption{\label{Fig:S2}ARPES spectra taken at $h\nu$ = 11.95 eV (10 K) which enhances the IP band in optimally doped Bi2223 ($T_{\rm{c}}$ = 110 K). (a) ARPES spectra in the nodal direction. (b) ARPES spectra in the off-nodal direction, and panel (b) near the Fermi level is magnified as shown in panel (c). These ARPES spectra cover the momentum space shown in Figs. 2 (a) and 2(b) of \cite{Kunisada} as shown in panel (d). Black and red markers indicate the peak positions of momentum-distribution curve and energy-distribution curve, respectively. }
\end{figure}

Since in synchrotron radiation ARPES, one can continuously change the photon energy, the electronic structure can be measured with varying matrix-element effects. When ARPES experiments were performed using 11.95 eV as shown in Fig. \ref{Fig:S2}, ARPES data in panels (b) and (c), which correspond to the same momentum space as in Fig. 2 of \cite{Kunisada} (reproduced in Fig. \ref{Fig:S2}(d)) show that the spectral intensity from the IP band is enhanced. According to these data, one can clearly see the band bending due to the SC gap in the present study. \\ 
\\

In a previous work \cite{kink}, we have reported that the nodal kink energy in the IP band of Bi2223 ($\sim$100 meV) is clearly different from that of Bi2212 ($\sim$70 meV), as shown in Fig. \ref{Fig:S2}(a) (red arrow). In the off-nodal direction shown in cut (2), the ARPES data show that the kink energy remains large in the IP band, and the difference between the band top and kink in energy of the IP band is estimated to be $\sim$55 meV as shown in Fig. \ref{Fig:S2}(c). Therefore, the SC gap ($\Delta_{\rm{IP}}$) of the IP band can be estimated from the IP band top as shown in Fig. \ref{Fig:S2}(c).

\subsection{\Large{III. Bare electronic structure}}
The electronic Hamiltonian for the triple-layer cuprate Bi2223, in the absence of 
electron-phonon coupling, is given by 
\begin{equation}
H_0 = \sum_{i,\bk,\sigma} \epsilon_i(\bk) d^\dagger_{i,\bk,\sigma}d^{\phantom{\dagger}}_{i,\bk,\sigma}
+\sum_{\bk,<i,j>,\sigma} [t_\perp(\bk) d^\dagger_{i,\bk,\sigma}d^{\phantom\dagger}_{j,\bk,\sigma} + H.C.] + 
\sum_{\bk,i} [\Delta_i(\bk)d^\dagger_{i,\bk,\uparrow}d^\dagger_{i,-\bk,\downarrow} + H.C.], 
\end{equation} 
where $d^\dagger_{i,\bk,\sigma}$ ($d^{\phantom\dagger}_{i,\bk,\sigma}$) creates (annihilates) 
an electron of spin $\sigma$ and momentum $\bk$ in plane $i$, $\epsilon_i(\bk)$ is the 
band dispersion of plane $i$ in the absence of interlayer hopping, $t_\perp(\bk)$ is the 
matrix element for single-particle hopping between neighboring layers, $\langle \dots\rangle$ 
denotes a sum over nearest neighbor planes, and $\Delta_i(\bk) = 
\Delta_{i}^0[\cos(k_xa/2) - \cos(k_y/2)]/2$ is the intralayer pairing potential in plane $i$. 
The bare band dispersions are given by 
\begin{equation}
\epsilon_i(\bk) = - 2t_i[\cos(k_xa)+\cos(k_ya)] - 4t^\prime_i\cos(k_xa)\cos(k_ya) 
- 2t^{\prime\prime}_i[\cos(2k_xa)+\cos(2k_ya)] - \mu_i,
\end{equation}
where we take (in units of eV) $t_{\rm{IP}} = 0.34$, $t^\prime_{\rm{IP}} = -0.26t_{\rm{IP}}$, 
$t^{\prime\prime} = -t^\prime_{\rm{IP}}/2$, and $\mu_{\rm{IP}} = -0.27$ for IP
and $t_{\rm{OP}} = 0.30$, $t^\prime_{\rm{OP}} = -0.29t_{\rm{OP}}$, $^{\prime\prime} = -t_{\rm{OP}}^\prime/2$, and $\mu_{\rm{OP}} = -0.33$ for OP. The hopping parameters are taken from [\onlinecite{tperp}] while the values $\mu_i$ have been adjusted to counter a small renormalization of the chemical potential. 
The interlayer hopping has the form appropriate for $c$-axis tunneling 
\begin{equation}
t_\perp(\bk) = t_{\perp}^{(0)}+t_\perp^{(1)}[\cos(k_xa) - \cos(k_ya)]^2/4, 
\end{equation} 
where we take $t_\perp^{(0)} = 0$ meV, $t_\perp^{(1)} = 56$ meV. Note that this value is significantly smaller than the 
 LDA estimate of 150 meV ([\onlinecite{Anderson}]).  
It was chosen to reproduce the data after the inclusion of the electron-phonon self-energy. 
We therefore interpret our $t_\perp{^{(1)}}$ value as an effective value which accounts for the additional self-energy 
effects.

\subsection{\Large{IV. Interlayer single-particle hopping}}
\begin{figure}[h] 
 \includegraphics[width=10cm]{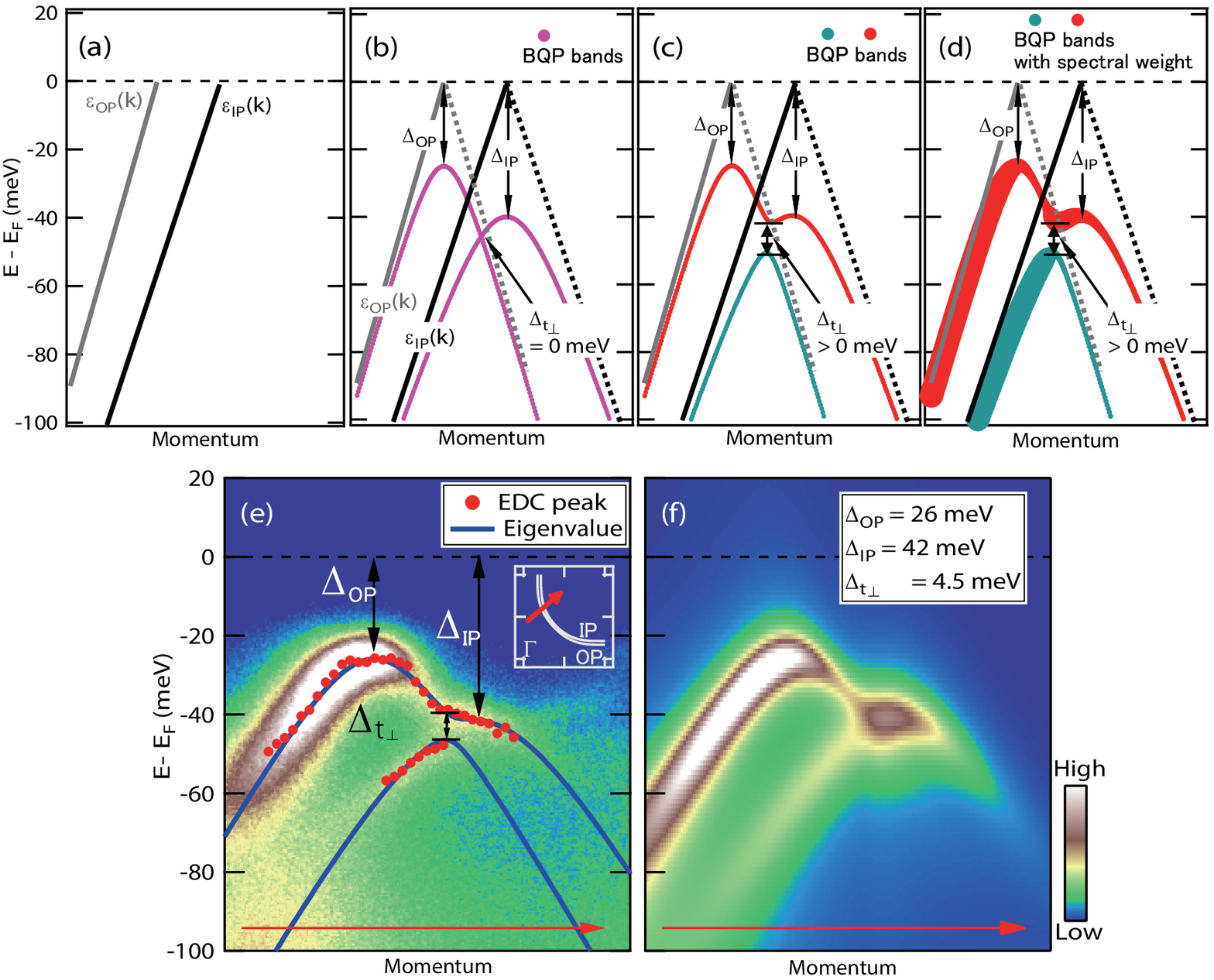}
 \caption{\label{Fig:S3}Schematic Bogoliubov quasi-particle (BQP) bands along an off-nodal cut in the first Brillouin zone.
(a): Bare bands for the OP and IP in Bi2223. 
(b): $\Delta_{\rm{OP}}$ and $\Delta_{\rm{IP}}$ have a finite value ($\Delta_{\rm{OP}}$ $<$ $\Delta_{\rm{IP}}$) but $\Delta_{{t}_{\perp}}$ is set at 0 meV. 
(c): Same as panel (b) but $\Delta_{{t}_{\perp}}$ has a finite value. 
(d) Same as panel (c) but the spectral weight deduced from Eq. (\ref{Eq:SPH_M}) is shown by symbol size.
(e): ARPES intensity plot in momentum space, cut$\#4$ in Fig. 1. EDC peak positions and eigenvalues deduced from Eq. (\ref{Eq:SPH}) are superimposed. 
(f): Reproduced intensity plots corresponding  to the ARPES spectra of panel (e). Parameters used in the calculation are shown in the inset. Note that, even if $\Delta_{t_{\perp}}$ is replaced by $\Delta_{\rm{OP-IP}}$ [Eq. (S9)], the fitted results do not change if an appropriate $\Delta_{\rm{OP-IP}}$ value is used.}
\end{figure}

If we assume the single-particle hopping $t_\perp$(\bk) between the IP and OP in the superconducting state, the Hamiltonian of the triple-layer cuprate becomes:
\begin{equation}\label{Eq:Mori}
{H} = 
\Phi^{\dagger}\left(
\begin{array}{cccccc}
\epsilon_{\rm{OP}}(\bk) & \Delta_{\rm{OP}}(\bk) & t_{\perp}(\bk) & 0 & 0 & 0\\
\Delta_{{\rm{OP}}}(\bk) & -\epsilon_{\rm{OP}}(\bk) &0 &  -t_{\perp}(\bk)  & 0 & 0\\
t_{\perp}(\bk) & 0 & \epsilon_{\rm{IP}}(\bk) & \Delta_{\rm{IP}}(\bk) & t_{\perp}(\bk) & 0\\
0 & -t_{\perp}(\bk) & \Delta_{\rm{IP}}(\bk) & -\epsilon_{\rm{IP}}(\bk) & 0 & -t_{\perp}(\bk)\\
0 & 0 & t_{\perp}(\bk) & 0 & \epsilon_{\rm{OP}}(\bk) & \Delta_{\rm{OP}}(\bk)\\
0 & 0 & 0 & -t_{\perp}(\bk) & \Delta_{\rm{OP}}(\bk) & -\epsilon_{\rm{OP}}(\bk) 
\end{array} 
\right)\Phi
\end{equation}
\begin{equation}\label{Eq:SPH_M}
=\Psi^\dagger\left(
\begin{array}{cccccc}
\epsilon_{\rm{OP}}(\bk) & \Delta_{\rm{OP}}(\bk) & \sqrt{2}t_{\perp}(\bk) & 0 & 0 & 0\\
\Delta_{{\rm{OP}}}(\bk) & -\epsilon_{\rm{OP}}(\bk) &0 &  -\sqrt{2}t_{\perp}(\bk)  & 0 & 0\\
\sqrt{2}t_{\perp}(\bk) & 0 & \epsilon_{\rm{IP}}(\bk) & \Delta_{\rm{IP}}(\bk) & 0 & 0\\
0 & -\sqrt{2}t_{\perp}(\bk) & \Delta_{\rm{IP}}(\bk) & -\epsilon_{\rm{IP}}(\bk) & 0 & 0\\
0 & 0 & 0 & 0 & \epsilon_{\rm{OP}}(\bk) & \Delta_{\rm{OP}}(\bk)\\
0 & 0 & 0 & 0 & \Delta_{\rm{OP}}(\bk) & -\epsilon_{\rm{OP}}(\bk) 
\end{array}
\right)\Psi,
\end{equation}
where 
\begin{equation}
\Phi\equiv \left(
\begin{array}{cccccc}
c_{k\uparrow\rm{OP1}}\\
c^{\dagger}_{-k\uparrow\rm{OP1}}\\
c_{k\uparrow\rm{IP}}\\
c^\dagger_{-k\uparrow\rm{IP}}\\
c_{k\uparrow\rm{OP2}}\\
c^{\dagger}_{-k\uparrow\rm{OP2}}
\end{array}
\right), 
\Psi\equiv \left(
\begin{array}{cccccc}
c_{k\uparrow\rm{BB}}\\
c^{\dagger}_{-k\uparrow\rm{BB}}\\
c_{k\uparrow\rm{IP}}\\
c^\dagger_{-k\uparrow\rm{IP}}\\
c_{k\uparrow\rm{AB}}\\
c^{\dagger}_{-k\uparrow\rm{AB}}
\end{array}
\right).\\
\end{equation}
$\Delta_{\rm{OP}}(\bk)$, $\Delta_{\rm{IP}}(\bk)$, $t_{\perp}(\bk)$, $\epsilon_{\rm{OP}}(\bk)$, and $\epsilon_{\rm{IP}}(\bk)$ are the superconducting gap of OP and IP, the single-particle hopping parameter, the bare OP and IP bands, respectively. 
From Eq. (S5), the eigenvalues of the IP-bonding OP hybridized bands are given by 
\begin{equation}\label{Eq:SPH}
E^2 = \frac{E_{\rm{IP}}^2+E_{\rm{OP}}^2}{2}+2t_{\perp}(\bk)^2 \pm \sqrt{\left(\frac{E_{\rm{IP}}^2-E_{\rm{OP}}^2}{2}\right)^2+2t_{\perp}(\bk)^2\{(\epsilon_{\rm{IP}}(\bk)+\epsilon_{\rm{OP}}(\bk))^2+(\Delta_{\rm{IP}}(\bk)-\Delta_{\rm{OP}}(\bk))^2}\},
\end{equation}
where $E_{\rm{IP}}^2=\epsilon_{\rm{IP}}(\bk)^2+\Delta_{\rm{IP}}(\bk)^2$ and $E_{\rm{OP}}^2=\epsilon_{\rm{OP}}(\bk)^2+\Delta_{\rm{OP}}(\bk)^2$, and the anti-crossing gap 2$\Delta_{{t}\perp}(\bk)$ obtained from the ARPES experiment in the present study is approximately expressed by 
\begin{equation}
2\Delta_{t_{\perp}}(\bk)\sim\frac{2\sqrt{2}t_{\perp}(\bk)|\Delta_{\rm{IP}}(\bk)-\Delta_{\rm{OP}}(\bk)|}{\sqrt{(\Delta_{\rm{IP}}(\bk)-\Delta_{\rm{OP}}(\bk))^2+(\epsilon_{\rm{OP}}(\bk)-\epsilon_{\rm{IP}}(\bk))^2}},
\end{equation}
where we assume that $t_{\perp}(\bf{k})$ is small. The experimental data are compared with the calculated results as shown in Fig. \ref{Fig:S3}.
\\
\newpage

\subsection{\Large{V. Interlayer Cooper pairing}}
One of the candidate mechanisms that enhance $T_{\rm{c}}$ in Bi2223 is the interlayer Cooper pairing between the IP and OP. 
In order to analyze the ARPES spectra assuming the interlayer Cooper pairing, we use the Hamiltonian:

\begin{equation}\label{Eq:CPH}
{H} = 
\left[
\begin{array}{cccc}
\epsilon_{\rm{OP}}(\bk) & \Delta_{\rm{OP}}(\bk) & 0 & \Delta_{\rm{OP-IP}}(\bk)\\
\Delta_{{\rm{OP}}}(\bk) & -\epsilon_{\rm{OP}}(\bk) & \Delta_{\rm{OP-IP}}(\bk) &0\\
0 & \Delta_{\rm{OP-IP}}(\bk) & \epsilon_{\rm{IP}}(\bk) &\Delta_{\rm{IP}}(\bk)\\
\Delta_{\rm{OP-IP}}(\bk) & 0 & \Delta_{\rm{IP}}(\bk) & -\epsilon_{\rm{IP}}(\bk)\\
\end{array} 
\right],
\end{equation}
From Eq. (\ref{Eq:CPH}), we also obtain the eigenvalues and the spectral weight of the IP and OP bands as the same as those in Fig. \ref{Fig:S3} calculated using Eq. (\ref{Eq:SPH_M}).

\subsection{\Large{VI. Hybridization of the BQP bands near the anti-nodal region}}

\begin{figure} [h]
 \includegraphics[width=\textwidth]{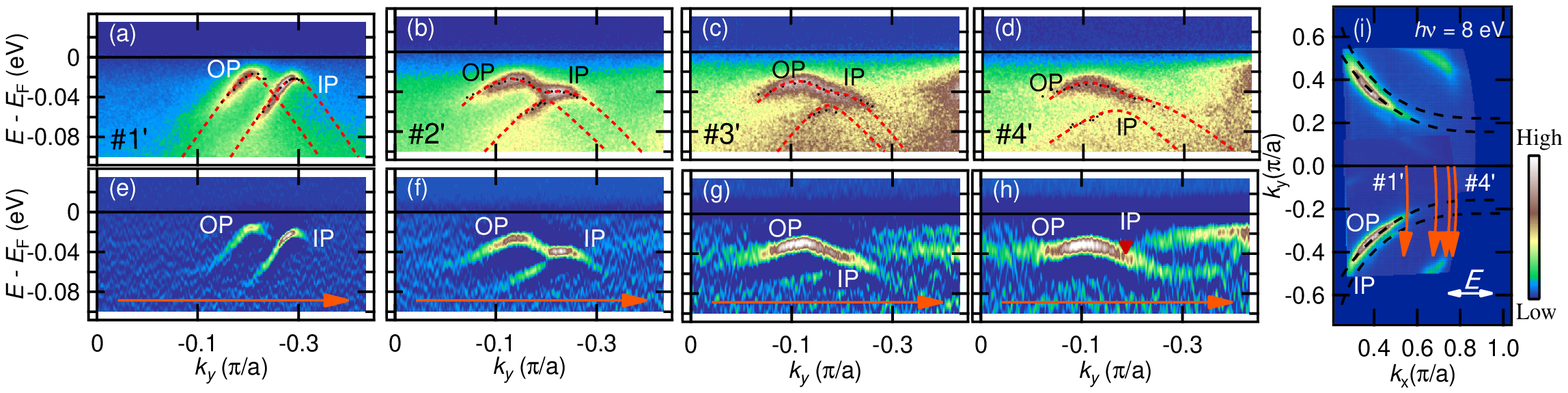}
 \caption{\label{Fig:HBQP} Hybridization of the BQP bands in Bi2223. (a)-(d) Raw energy-momentum plots corresponding to several momentum cuts ($\#$1$^{\prime}$ - $\#$4$^{\prime}$) shown in the Fermi surface (FS) mapping (i). EDC peak positions are displayed by black marker as a guide to the eye and fitted by using Eq. (S7) (red dotted lines). Parameters of fitting are obtained from the results of Fig. 2(j). (e)-(h) Second-derivatives with respect to energy of the intensity plots (a)-(d).}
\end{figure}

In Figs. \ref{Fig:HBQP}(a)-\ref{Fig:HBQP}(h), we plot data for momentum cuts closer to the anti-node. The hole branch of the OP band is more clearly seen due to the stronger hybridization with the IP band owing to the larger $\Delta_{t_{\perp}}$($k$) than that near the node. The anti-crossing gap closes above $T_{\rm{c}}$ as shown in Fig. \ref{tempdep}.

\subsection{\Large{VII. Self-energy due to electron-phonon interactions}}
The Hamiltonian for the electron-phonon interaction is of the form
\begin{equation}
H_{e-ph} = -\frac{1}{\sqrt{N}}\sum_{\bk,\bq,\sigma,\nu,i} 
g_{i,\nu}(\bk,\bk-\bq) d^\dagger_{i,\bk,\sigma}d^{\phantom\dagger}_{i,\bk-\bq,\sigma}
(b^\dagger_{i,\bq,\nu} + b^{\phantom\dagger}_{i,-\bq,\nu}).
\end{equation}
Here $b^\dagger_{i,\bq,\nu}$ ($b^{\phantom\dagger}_{i,\bq,\nu}$) 
creates (annihilates) a phonon mode of momentum $\bq$ in branch $\nu$ of plane $i$, 
and $g_{i,\nu}(\bk,\bk-\bq)$ is the electron-phonon coupling constant.    
We include coupling a spectrum of oxygen branches; this includes the $c$-axis 
polarized out-of-phase bond-buckling (the so-called $B_{1g}$ modes) and  
in-plane polarized bond-stretching 
(the so-called breathing modes) optical modes \cite{Aspects,ScreeningPRB}, as well as the in-plane transverse acoustic modes \cite{AcousticPRL}. 
The optical modes are assumed to be dispersionless ($\Omega_{b1g} = 36$ meV, 
$\Omega_{br} = 70$ meV) 
and the momentum dependence of the coupling constants are given by in [\onlinecite{ScreeningPRB}] [Eqs. (16) and (17) for the breathing and $B_{1g}$ 
modes, respectively].  
The coupling constant for the acoustic branch 
is treated as outlined in [\onlinecite{AcousticPRL}], where we assume 
$q_{TF} = 0.6/a$ for the outer CuO$_2$ planes and $q_{TF} = 0.4/a$ for the 
inner CuO$_2$ plane. 

In the Nambu notation, the self-energy for band $i = $ IP, OP is partitioned 
in the usual way
\begin{equation}
\hat{\Sigma}_i(\bk,\omega) = \omega[1 - Z_i(\bk,\omega)]\hat{\tau}_0 + 
\chi_i(\bk,\omega)\hat{\tau}_3 + \phi_i(\bk,\omega)\hat{\tau}_1, 
\end{equation}
where $\hat{\tau}_\alpha$ are the usual Pauli matrices. 
The electron-phonon self-energies are calculated assuming a single iteration of the Eliashberg equations, 
as outlined in a number of previous works \cite{ScreeningPRB, Aspects, AcousticPRL}. For simplicity, 
we neglect any interlayer tunneling $t_\perp(\bk)$ and interplane pairing potential when 
calculating the electron-phonon contribution to the self-energy. The imaginary part of the 
self-energies for plane $i = $ IP, OP are given by \cite{Aspects,AcousticPRL}
\begin{eqnarray}\label{Eq:Z}\nonumber
\omega Z^{\prime\prime}_{i}(\bk,\omega)&=&\frac{\pi}{2N}\sum_{\bp,\nu} |g_{i,\nu}(\bk,\bq)|^2 
   \big( [n_b(\Omega_{\nu,\bq}) + n_f(E_{i,\bp})]
          [\delta(\omega+\Omega_{\nu,\bq} - E_{i,\bp}) + 
           \delta(\omega-\Omega_{\nu,\bq} + E_{i,\bp})] \\ & & + 
          [n_b(\Omega_{\nu,\bq}) + n_f(-E_{i,\bp})]
          [\delta(\omega-\Omega_{\nu,\bq} - E_{i,\bp}) + 
           \delta(\omega+\Omega_{\nu,\bq} + E_{i,\bp})] \big)
\end{eqnarray}
\begin{figure} [b]
 \includegraphics[width=\textwidth]{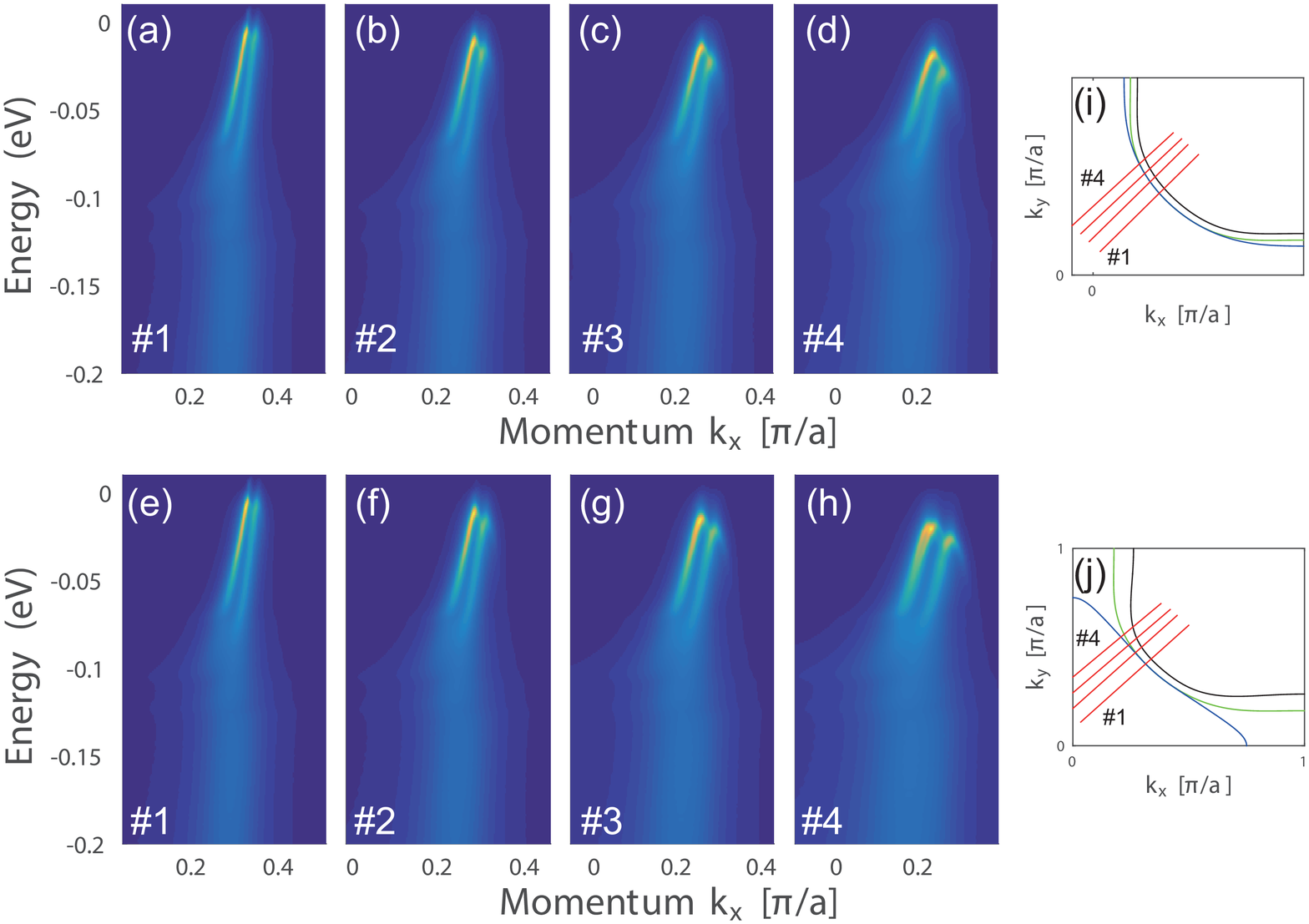}
 \caption{\label{Fig:Akw} Calculated spectral function along four cuts in the 
 first Brillouin zone moving from near nodal (left) to near anti-nodal (right) (a)-(d): Result using $t_{\perp}^{(1)}$ = 56 meV. (e)-(h): Result using $t_{\perp}^{(1)}$ = 200 meV. (i) and (j) are the calculated Fermi surfaces for $t_{\perp}^{(1)}$ = 56 meV and 200 meV, respectively. The value of $t_{\perp}$ contributes to the splitting between the bonding and anti-bonding bands but, in the present model, the SC gaps $\Delta_{\rm{OP}}$ and $\Delta_{\rm{IP}}$ do not depend on $t_{\perp}$.}
\end{figure}
\begin{eqnarray}\label{Eq:Chi}\nonumber
 \chi^{\prime\prime}_{i}(\bk,\omega)&=&-\frac{\pi}{2N}\sum_{\bp,\nu} |g_{i,\nu}(\bk,\bq)|^2 
   \frac{\epsilon_{i,\bp}}{E_{i,\bp}}
   \big( [n_b(\Omega_{\nu,\bq}) + n_f(E_{i,\bp})]
          [\delta(\omega+\Omega_{\nu,\bq} - E_{i,\bp}) - 
           \delta(\omega-\Omega_{\nu,\bq} + E_{i,\bp})] \\ & & + 
          [n_b(\Omega_{\nu,\bq}) + n_f(-E_{i,\bp})]
          [\delta(\omega-\Omega_{\nu,\bq} - E_{i,\bp}) - 
           \delta(\omega+\Omega_{\nu,\bq} + E_{i,\bp})] \big)
\end{eqnarray}

\begin{eqnarray}\label{Eq:Phi}\nonumber
 \phi^{\prime\prime}_{i}(\bk,\omega)&=&\frac{\pi}{2N}\sum_{\bp,\nu} |g_{i,\nu}(\bk,\bq)|^2 
   \frac{\Delta_{i,\bp}}{E_{i,\bp}}
   \big( [n_b(\Omega_{\nu,\bq}) + n_f(E_{i,\bp})]
          [\delta(\omega+\Omega_{\nu,\bq} - E_{i,\bp}) - 
           \delta(\omega-\Omega_{\nu,\bq} + E_{i,\bp})] \\ & & + 
          [n_b(\Omega_{\nu,\bq}) + n_f(-E_\bp)]
          [\delta(\omega-\Omega_{\nu,\bq} - E_{i,\bp}) - 
           \delta(\omega+\Omega_{\nu,\bq} + E_{i,\bp})] \big)
\end{eqnarray}
where $n_f$ and $n_b$ are the usual Fermi and Bose occupation numbers, respectively, and 
$E_{i,\bk}^2 = \epsilon_{i,\bk}^2 + \Delta_{i,\bk}^2$. 
The real parts of the self-energies are obtained from the 
Kramers-Kronig relations.

Once the self-energies have been calculated, the interlayer tunneling and pairing 
potential are introduced by hand and the full $6\times 6$ matrix Green's 
function $\hat{G}(\bk,\omega)$ is formed 
\begin{equation}\label{Eq:BigG}
\hat{G}^{-1}(\bk,\omega) = 
\left[
\begin{array}{ccc}
\hat{I}(\omega+i\delta) - \hat{\Sigma}_{OP}(\bk,\omega)  & 0 & -t_\perp(\bk)\hat{\tau}_3  \\
0 & \hat{I}(\omega+i\delta) - \hat{\Sigma}_{OP}(\bk,\omega) & -t_\perp(\bk)\hat{\tau}_3 \\
-t_\perp(\bk)\hat{\tau}_3 & -t_\perp(\bk)\hat{\tau}_3 & \hat{I}(\omega+i\delta) - \hat{\Sigma}_{IP}(\bk,\omega) 
\end{array} 
\right], 
\end{equation}
where each element of the matrix in Eq. (\ref{Eq:BigG}) is understood to be a $2\times 2$ matrix, and   
$\hat{I}$ is the identity matrix. The spectral function shown in 
Fig. \ref{Fig:Akw} is then obtained by inverting Eq. (\ref{Eq:BigG}) and tracing its imaginary 
part in the electron channel
\begin{equation}
A(\bk,\omega) = -\frac{1}{\pi} \sum_{i=1,3} \mathrm{Im}[G(\bk,\omega)]_{2i-1,2i-1}.
\end{equation}
The calculated spectral function for $t_{\perp}^{(1)}$ = 56 meV and $t_{\perp}^{(1)}$ = 200 meV is shown in Fig. \ref{Fig:Akw} for several cuts in the 
first Brillouin zone.  
\newpage
\subsection{\Large{VIII. Temperature dependence of the anti-crossing gap $\Delta_{t_{\perp}}$}}
\begin{figure} [h]
 \includegraphics[width=15cm]{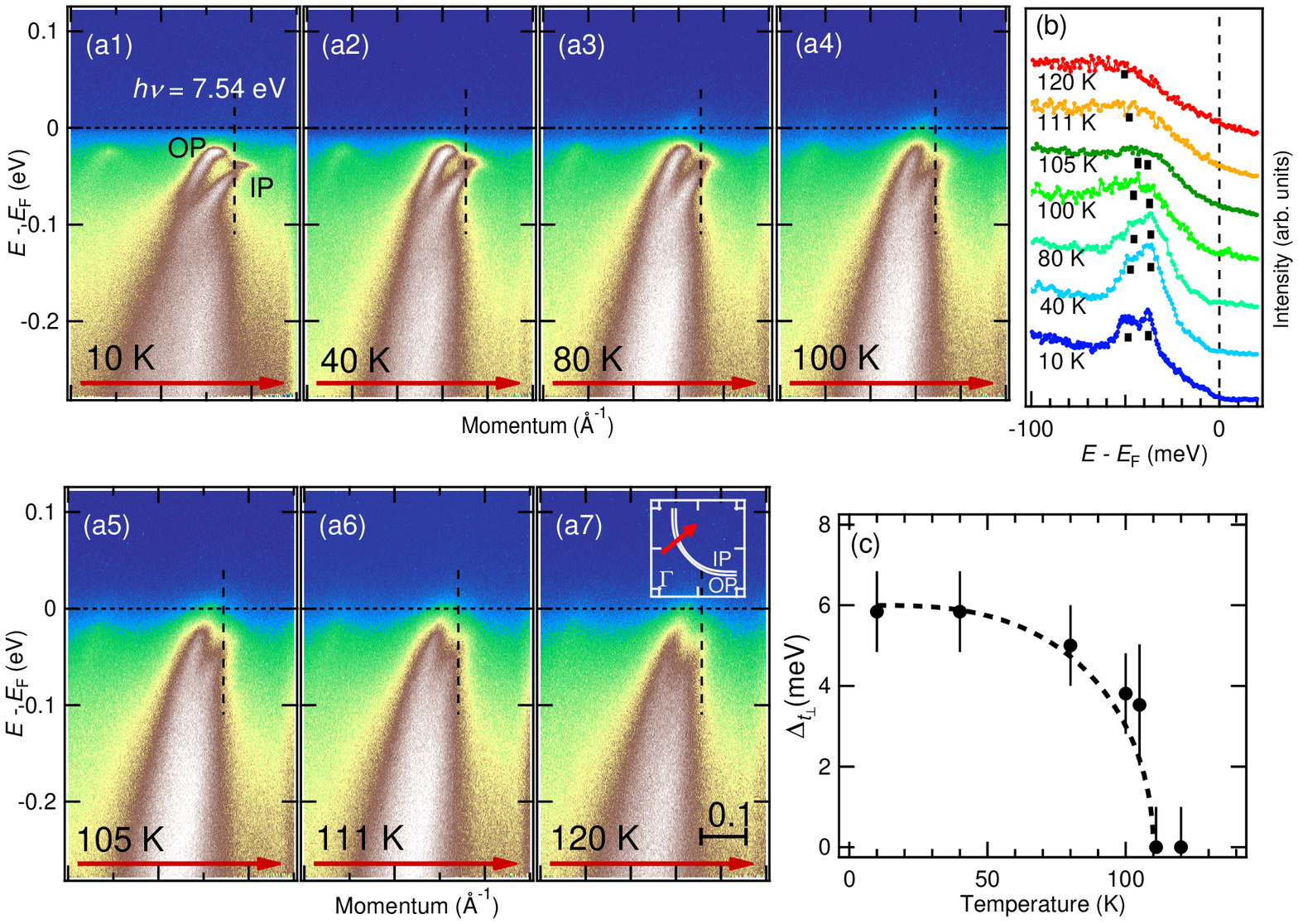}
 \caption{\label{tempdep} Temperature dependence of the anti-crossing gap in the off-nodal region of Bi2223 ($T_{\rm{c}}$ = 110 K). (a1)-(a7): ARPES intensity plots taken at various temperatures. (b): EDCs extracted from the ARPES intensity plot at momentum indicated by vertical dashed lines in panels (a1)-(a7). Vertical bars indicate the peak positions of the EDC.  (c) Temperature dependence of the anti-crossing gap $\Delta_{t_{\perp}}$ for cut $\#$4 in Figs. 1 and 2. The broken curve represents BCS-like temperature dependence.}
\end{figure}

We have measured the temperature dependence of ARPES spectra across $T_{\rm{c}}$, and confirmed that $\Delta_{t_{\perp}}$ indeed closes above $T_{\rm{c}}$, consistent with the previous study \cite{Kunisada}. The temperature dependence of ARPES spectra for cut $\#$4 of Fig. 1 is shown in Fig. \ref{tempdep}. As the temperature increases, the band dispersions and the BQP hybridization becomes blurred. The EDC at the $k_{\rm{F}}$ of the IP band shows two peaks separated by the anti-crossing gap 2$\Delta_{t_{\perp}}$ and the gap gradually closes towards $T_{\rm{c}}$ = 110 K, resulting in a BCS-like gap closure. The temperature evolution of the EDCs thus strongly suggests that the anti-crossing gap 2$\Delta_{t_{\perp}}$ is indeed intimately related to the occurrence of superconductivity, consistent with the previous work. 

\newpage
\subsection{\Large{IX. Estimation of the $T_{\rm{c}}$ enhancement in triple-layer cuprates}}
We can estimate the $T_{\rm{c}}$ enhancement due to the electron-phonon coupling by adopting a multi-channel square-well model
similar to that used in [4]. Here we assume contributions from four bosonic modes of frequency $\Omega_{\nu}$ ( $\nu$=
1, 2, 3, and 4) and projected couplings $\lambda_{z,\nu}$ and $\lambda_{\phi,\nu}$. The modes are associated with the acoustic phonon branch
($\Omega_1$ = 10 meV), the $c$-axis bond Cu-O buckling branch ($\Omega_2$ = 36 meV), the in-plane bond stretching breathing branch
($\Omega_3$ = 70 meV), and finally a spectrum of high-energy spin fluctuations ($\Omega_4\sim$ 2$J$ $\sim$ 260 meV). Following standard
approximations, we arrive at the set of coupled equations ($\hbar$ = $k_b$ = 1)

\begin{eqnarray}\label{Eq:four-well-model}\nonumber
Z_1\Delta_1 = \tilde{\lambda}_1\Delta_1F(0,\Omega_1)+\tilde{\lambda}_1\Delta_2F(\Omega_1,\Omega_2)+\tilde{\lambda}_1\Delta_3F(\Omega_2,\Omega_3)+\tilde{\lambda}_1\Delta_4F(\Omega_3,\Omega_4)\nonumber\\
Z_2\Delta_2 = \tilde{\lambda}_2\Delta_1F(0,\Omega_1)+\tilde{\lambda}_2\Delta_2F(\Omega_1,\Omega_2)+\tilde{\lambda}_2\Delta_3F(\Omega_2,\Omega_3)+\tilde{\lambda}_2\Delta_4F(\Omega_3,\Omega_4)\nonumber\\
Z_3\Delta_3 = \tilde{\lambda}_3\Delta_1F(0,\Omega_1)+\tilde{\lambda}_3\Delta_2F(\Omega_1,\Omega_2)+\tilde{\lambda}_3\Delta_3F(\Omega_2,\Omega_3)+\tilde{\lambda}_3\Delta_4F(\Omega_3,\Omega_4)\\
Z_4\Delta_4 = \tilde{\lambda}_4\Delta_1F(0,\Omega_1)+\tilde{\lambda}_4\Delta_2F(\Omega_1,\Omega_2)+\tilde{\lambda}_4\Delta_3F(\Omega_2,\Omega_3)+\tilde{\lambda}_4\Delta_4F(\Omega_3,\Omega_4)\nonumber
\end{eqnarray}
where $Z_i$ = $1 + \sum^4_{j=i} \lambda_{z,j}, \tilde{\lambda_i} = \sum^4_{j=i} \lambda_{\phi,j}$, and $F$($\Omega_1, \Omega_2$) = $\int_{\Omega_1}^{\Omega_2}$ $dz/z$ $\tanh(z/2T)$. This system can be solved analytically for non-zero solutions of $\Delta_i$ at $T$ = $T_c$. The final result is
\begin{eqnarray}\label{Eq:Tc}
T_c = 1.134\Omega_1^{1-\alpha}\Omega_2^{\alpha-\beta}\Omega_3^{\beta-\gamma}\Omega_4^{\gamma}\exp\left({-\frac{1+\lambda_{z,1}+\lambda_{z,2}+\lambda_{z,3}+\lambda_{z,4}}{\lambda_{\phi,1}+\lambda_{\phi,2}+\lambda_{\phi,3}+\lambda_{\phi,4}}}\right)
\end{eqnarray}

The simulated spectral functions produce the following estimates for the optical mode's $\lambda$ values: The $B_{1g}$ branch in
the IP/OP: $\lambda_z$ = 0.31, $\lambda_{\phi}$ = 0.16 / $\lambda_z$ = 0.89, $\lambda_{\phi}$ = 0.56. The breathing branch in the IP/OP: $\lambda_z$ = 0.62, $\lambda_{\phi}$ = -0.25 / $\lambda_z$ = 0.93, $\lambda_{\phi}$ = -0.43. The negative sign for $\lambda_{\phi}$ indicates that the breathing modes are pair breaking. We can proceed to a $T_{\rm{c}}$ estimate in a few ways (we stress here that these are very rough estimates). In Fig. 4,  $T_{\rm{c}}$ of the main text is plotted as a function of the acoustic mode's contribution for two cases. Here, the  values for the two optical branches have been set to their average value in the inner and outer planes. As can be seen in Fig. 4(a), the pair-breaking contribution from the breathing modes significantly suppresses the $T_{\rm{c}}$ provided by the spin fulctuations. However, it should be noted here that the breathing modes should be heavily screened, and this has not been accounted for here. In Fig. 4(b), the pair breaking contributions from the breathing modes have been neglected under the assumption that they are well screened. The $B_{1g}$ modes give a significant enhancement to $T_{\rm{c}}$ but this has the effect of reducing the boost provided by the acoustic modes, making them slightly detrimental in the strong coupling limit (this arises due to increases in the quasiparticle mass).

\end{document}